\documentclass[aps,prl,twocolumn,superscriptaddress,showpacs]{revtex4}
\usepackage{psfig}
\usepackage{epsfig}
\usepackage{color}
\newcommand{\beq}{\begin{equation}}
\newcommand{\eeq}{\end{equation}}
\newcommand{\beqa}{\begin{eqnarray}}
\newcommand{\eeqa}{\end{eqnarray}}
\newcommand{\ba}{\begin{array}}
\newcommand{\ea}{\end{array}}

\begin{document}

\widetext 
\title{$1/f^{\alpha}$ noise in spectral fluctuations of quantum systems} 

\author{J.M.G. G\'omez} 
\author{A. Rela\~no} 
\author{J. Retamosa} 
\affiliation{Departamento de F\'{\i}sica At\'omica, Molecular y Nuclear, 
Universidad Complutense de Madrid, E-28040 Madrid, Spain} 

\author{E. Faleiro} 
\affiliation{Departamento de F\'{\i}sica Aplicada, EUIT Industrial,  
Universidad Polit\'ecnica de Madrid, 
E-28012 Madrid, Spain}

\author{L. Salasnich} 
\affiliation{INFM, Unit\`a di Milano Universit\`a, Dipartimento di
Fisica, Universit\`a di Milano, Via Celoria 16, I-20133 Milano, Italy}
\affiliation{CAMTP, University of Maribor, Krekova 2, 
SI-2000 Maribor, Slovenia}

\author{M. Vrani\v car} 
\author{M. Robnik} 
\affiliation{CAMTP, University of Maribor, 
Krekova 2, SI-2000 Maribor, Slovenia} 

\begin{abstract} 
The power law $1/f^{\alpha}$ in the power spectrum characterizes the
fluctuating observables of many complex natural systems.  Considering
the energy levels of a quantum system as a discrete time series where
the energy plays the role of time, the level fluctuations can be
characterized by the power spectrum. Using a family of quantum
billiards, we analyze the order to chaos transition in terms of this
power spectrum.  A power law $1/f^{\alpha}$ is found at all the
transition stages, and it is shown that the exponent $\alpha$ is
related to the chaotic component of the classical phase space of the
quantum system. 
\end{abstract} 


\maketitle

Of the different features which characterize complex physical systems,
perhaps the most ubiquitous, interesting and puzzling is the presence
of $1/f^{\alpha}$ noise \cite{Mandelbrot:99} in fluctuating physical
variables, i.e. the Fourier power spectrum $S(f)$ behaves as
$1/f^{\alpha}$ in terms of the frequency $f$. This kind of noise has
been detected in condensed matter systems, traffic engineering, DNA
sequence, quasar emissions, river discharge, human behavior,
heartbeat and dynamic images, among many others. Despite this
ubiquity, there is no universal explanation about this phenomenon. It
does not arise as a consequence of particular physical interactions,
but it is a generic manifestation of complex systems.

Recently, it was conjectured that {\it the energy spectra of chaotic
quantum systems are characterized by 1/f noise} \cite{Relano:02} . The
original idea was that the sequence of discrete energy levels in a
quantum system can be considered as a discrete time series, where the
energy plays the role of time. In that case, the energy level
fluctuations can be studied using traditional methods of time series
analysis, like the study of the power spectrum. When the idea was
applied to typical chaotic quantum systems, the power spectrum showed
a very accurate $1/f$ behavior \cite{Relano:02}. Hence chaotic quantum
systems can be added to the long list of complex natural systems which
exhibit $1/f$ noise. However, this new point of view also raises new
questions.  Is this a consequence of the universal behavior of
fluctuations in chaotic quantum systems? What happens in quantum
systems which are neither fully chaotic nor fully regular? In this
paper we try to find some answers using a quantum billiard to study
the power spectrum in the order--to--chaos transition. As shown below,
the ubiquitous $1/f^{\alpha}$ noise appears at all the transition
stages, with the exponent smoothly decreasing from ${\alpha}=2$ in a
regular system to ${\alpha}=1$ in a chaotic system. This is quite a
remarkable result indeed, since it contradicts the predictions of the
strict semiclassical limit \cite{Robnik:03}.

The concept of quantum chaos, or wave chaos in more general terms
\cite{Stockmann:99}, has no unique precise definition as yet, but
definitely can be described as quantum or wave like signatures of
classical chaos. It is well known that there is a relationship between
the energy level fluctuation properties of a quantum system and the
dynamics of its classical limit. Classically integrable systems give
rise to uncorrelated adjacent energy levels, which are well described
by Poisson statistics \cite{Berry:77}. In contrast, spectral
fluctuations of a quantum system whose classical limit is fully
chaotic (ergodic) show a strong repulsion between energy levels and
follow the predictions of random matrix theory (RMT) \cite{Bohigas:84,
Casati:80}. In practice, quantum systems without classical limit are
assumed to be chaotic when their fluctuations coincide with RMT
predictions.

The essential feature of chaotic energy spectra is the existence of
level repulsion and correlations (leading to strong spectral
rigidity), i. e. the spacing of two adjacent levels is unlikely to
deviate much from the mean spacing. This property is similar to the
{\it antipersistence} characteristic of some time series
\cite{Mandelbrot:99}.  Antipersistence, with different intensity
degrees, appears in time series with $1/f^{\alpha}$ noise, with
$1<\alpha<2$. Could the analogue level repulsion feature be also
associated to $1/f^{\alpha}$ noise?

To study the spectral fluctuations of quantum systems we follow the
method introduced in \cite{Relano:02}.  We use the statistic
$\delta_n$ defined by
 
\beq 
\delta_n =
\sum_{i=1}^n\left(s_i-\left<s\right>\right)=\epsilon_{n+1}-\epsilon_1-n,
\eeq
where $\epsilon_i$ are the unfolded energy levels,
$s_i=\epsilon_{i+1}-\epsilon_i$, and $<s>=1$ is the average value of
$s_i$. Thus $\delta_n$ represents the fluctuation of the $n$th excited
state.  Formally $\delta_n$ is similar to a time series where the
level order index $n$ plays the role of a discrete time. Therefore the
statistical behavior of level fluctuations can be investigated
studying the power spectrum $S(k)$ of the signal, given by

\beq 
S(k)=\left| {1\over \sqrt{M}} \sum_{n=1}^M \delta_n \exp{\left(
{-2\pi i k n\over M} \right)} \right|^2 \; , 
\eeq 
where $M$ is the size of the series and $f=2{\pi} k/M$ plays the role
of a frequency.

To investigate the behavior of $S(k)$ in the mixed regime between
integrability and chaos, we analyze it in the Robnik billiard
\cite{Robnik:83}.  Quantum billiards are considered as a paradigm in
quantum chaos. They have a discrete spectrum with an infinite number
of eigenvalues, and therefore it is possible to reach high statistical
precision by computing a large number of them.  Furthermore, they can
also be studied experimentally \cite{Stockmann:99,Veble:00}.

The boundary of the Robnik billiard is defined as the set of points
$w$ in the complex plane ${\bf C}$ which satisfy the equation
$w=z+\lambda z^2$, where $|z|=1$ and $\lambda$ is the deformation
parameter. It has been shown \cite{Robnik:83} that this billiard
exhibits a smooth transition from the integrable case ($\lambda =0$)
to an almost chaotic case ($1/4\leq \lambda \leq 1/2$). In order to
obtain a smooth analytic boundary, $\lambda$ must lie in the interval
$[0,1/2)$.  The Robnik billiard is one of the best systems to
investigate the order--to--chaos transition \cite{Robnik:83,
Robnik:84, Prosen:93}.  Compared to other quantum billiards, it has
the advantage that there are no bouncing ball orbits. For small values
of $\lambda$ the billiard is a typical KAM system, whereas for larger
values of $\lambda$ only one chaotic region dominates the phase space
with only few stability islands covered with invariant tori. The total
area in the bounce map (Poincar\'e surface of section) of these
invariant tori decreases monotonically with $\lambda$ and becomes
negligible when the shape of the billiard becomes non-convex, for
$\lambda > 1/4$. For $\lambda =1/2$ it has been shown rigorously by
Markarian \cite{Markarian:93} that the billiard is ergodic.

\begin{figure}
\centerline{\psfig{file=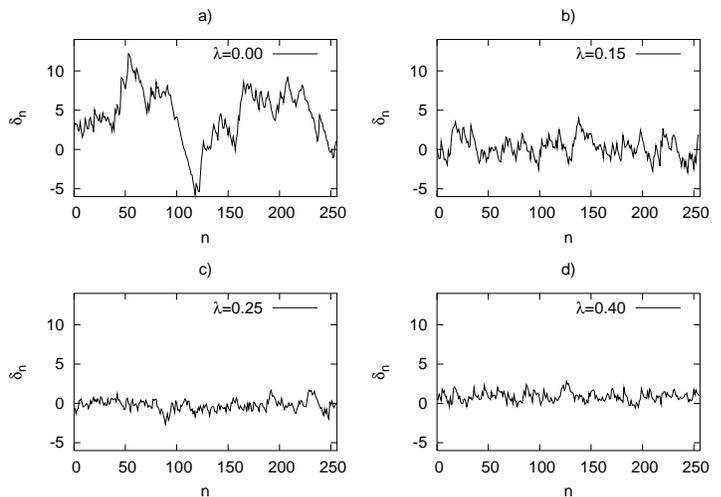,height=3.8in,angle=-90}}
\caption{Plot of the statistic $\delta_n$ for a set of 256 consecutive
energy levels of odd parity in the Robnik billiard, for several values
of the deformation parameter $\lambda$.}
\label{fig:delta_n}
\end{figure} 

The quantum energy levels $E_n$ of the Robnik billiard are numerically
calculated by solving the stationary Schr\"odinger equation of a free
particle whose wave function $\psi(w)$ is zero at the boundary of the
billiard. The billiard has reflection symmetry with respect to the
real axis, so there are two types of states: those with even parity
$\psi(w)=\psi(w^*)$ and those with odd parity $\psi(w)=-\psi(w^*)$;
odd and even parity states must be treated separately
\cite{Robnik:84, Prosen:93}. For each symmetry, our calculation
uses approximately 80,000 basis states, giving good eigenvalues for
about 65,000 levels for $\lambda=0$ and about 30,000 levels for
$\lambda=0.5$.

Fig. \ref{fig:delta_n} shows the energy level fluctuations of the
Robnik billiard given by $\delta_n$. It illustrates the effect of
level repulsion in the order--to--chaos transition, and its
relationship with the antipersistence of $\delta_n$ considered as a
time series. For the regular system ($\lambda=0$), the levels are
uncorrelated and therefore $\delta_n$ is neither persistent nor
antipersistent. As $\lambda$ increases the system becomes more
chaotic, and $\delta_n$ looks like a typical antipersistent series in
the almost chaotic region for $\lambda>1/4$.

\begin{figure}[h]
\centerline{\psfig{file=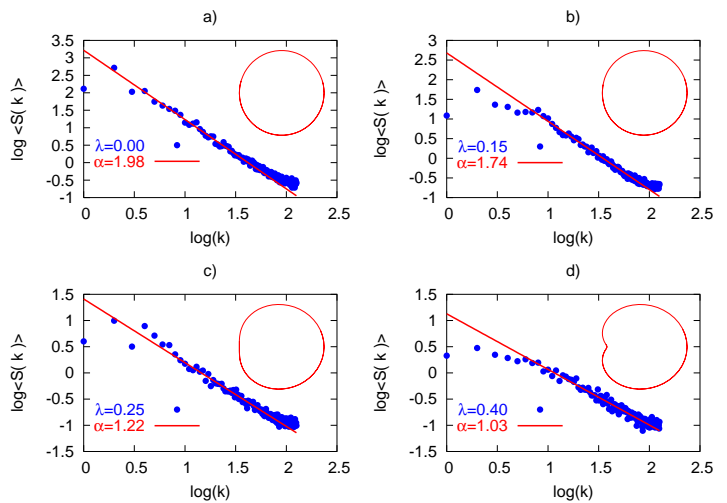,height=3.8in,angle=-90}}
\caption{Average power spectrum $\langle S(k) \rangle$ of the
statistic $\delta_n$ for the odd parity energy levels corresponding to
the shapes of the Robnik billiard inserted in the figures. The four
values of the deformation parameter $\lambda$ are the same as in
Fig. \ref{fig:delta_n}.  The solid (red) line is the best fit to the
power law $1/k^{\alpha}$.}
\label{fig:s_k}
\end{figure}

Fig. \ref{fig:s_k} shows the power spectrum of $\delta_n$ obtained
from the energy levels $E_n$ of the Robnik billiard for several values
of $\lambda$. The contour of the billiard is shown as an inset in each
panel. We calculate an ensemble average of $S(k)$ in order to reduce
statistical fluctuations and clarify its main trend. The average
$\langle S(k)\rangle$ is calculated with $25$ sets of $256$
consecutive high energy levels of odd parity. It is clearly seen that
for each value of $\lambda$ it follows the scaling law
\beq 
\langle S(k) \rangle \sim {1\over k^{\alpha}} \; , 
\eeq
where $\alpha$ depends on $\lambda$. In fact the fit of $\langle
S(k)\rangle $ to the power law $1/k^{\alpha}$ is excellent. In all
cases the error in the linear regression is less than $3\%$.  For
$\lambda =0$ (integrable case) the exponent is $\alpha = 1.98$, as
expected for uncorrelated energy levels. As $\lambda$ increases the
exponent $\alpha$ decreases and becomes $\alpha \simeq 1 $ for
$\lambda \simeq 1/2$.  Thus, $\alpha$ may serve as a measure of the
chaoticity of the system, since it changes from $\alpha=2$ for regular
systems to $\alpha=1$ for chaotic ones.

\begin{figure}[h]
\centerline{\psfig{file=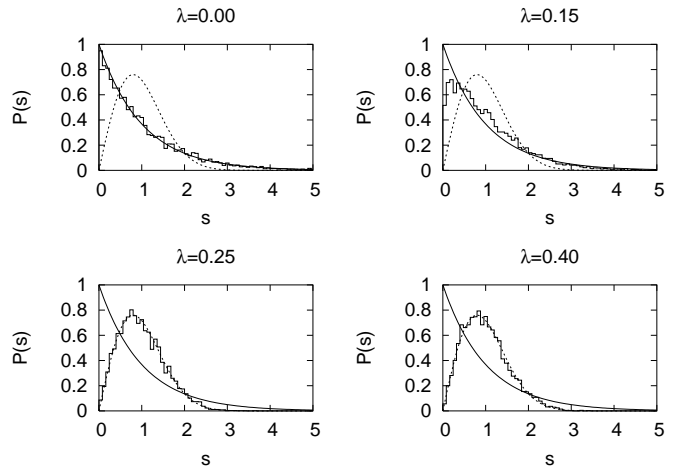,height=3.6in,angle=-90}}
\caption{Nearest neighbor spacing distributions for the spectra of
four different shapes of the Robnik billiard. The solid line is the
Poisson distribution and the dotted line corresponds to the Wigner
distribution predicted by RMT.}
\label{fig:nnsd}
\end{figure}

\begin{figure}[h]
\centerline{\psfig{file=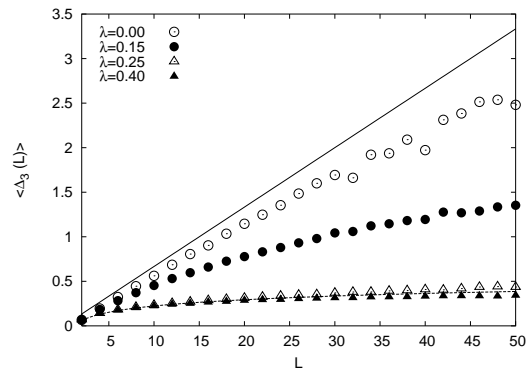,height=2.8in,angle=-90}}
\caption{Spectral average $\langle\Delta_3(L)\rangle$ values
calculated numerically for the Robnik billiard using the same
deformation parameters as in previous figures. The results of the
Poisson statistics (solid line) and RMT (dotted line) are also
displayed.}
\label{fig:delta3}
\end{figure}

It is worth to compare this result with the behavior of more
conventional statistics, like the nearest neighbor spacing
distribution $P(s)$, which measures short range correlations, and the
Dyson $\Delta_3(L)$ statistic appropriate for correlations of length
$L$ \cite{Mehta:91}. Fig. \ref{fig:nnsd} displays $P(s)$ for several
values of $\lambda$. At $\lambda=0$ the histogram follows the
predicted curve for regular systems (Poisson limit). For
$\lambda=0.15$, $P(s)$ deviates from Poisson toward the RMT limit. As
we shall see below, this behavior of $P(s)$ reflects that the
underlying classical dynamics is neither regular nor ergodic
(chaotic). Finally, for $\lambda=0.25$ and $\lambda=0.4$ the system
exhibits short range correlations characteristic of chaotic systems
(RMT limit). Fig. \ref{fig:delta3} shows the spectral average
$\langle\Delta_3(L)\rangle$ for energy intervals of length $L$ ranging
from $L=2$ to $L=50$, for several values of $\lambda$. The spectral
average is calculated using $25$ sets of $L$ consecutive high energy
levels to avoid, as far as possible, the influence of short periodic
orbits. The evolution of this statistic with $\lambda$ is analogous to
that of $P(s)$.  When $\lambda=0$, $\left<\Delta(L)\right>$ falls near
the Poisson prediction for regular systems, and for $\lambda=0.25$ and
$0.4$ it is almost indistinguishable from the RMT prediction for
chaotic systems. Therefore, $P(s)$ and $\langle\Delta_3(L)\rangle$
have a smooth behavior in terms of $\lambda$. Nevertheless, they move
faster than $\delta_n$ toward the RMT limit as $\lambda$
increases. For instance, both $P(s)$ and $\langle\Delta_3(L)\rangle$
coincide with RMT predictions for $\lambda=0.25$, while $\delta_n$
still points to an intermediate regime between regularity and chaos.

\begin{figure}[h]
\centerline{\psfig{file=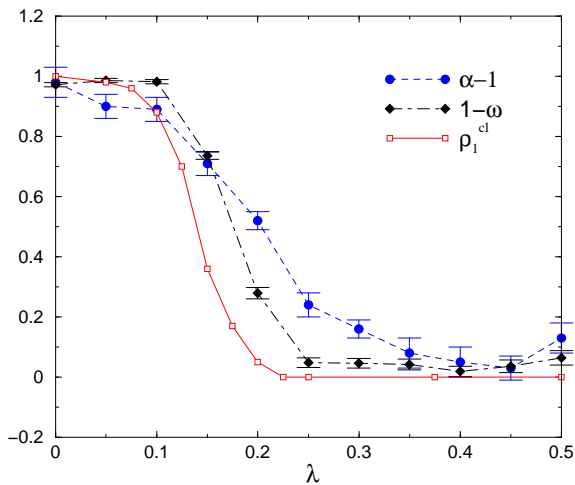,height=2.6in}} 
\caption{Behavior of the power spectrum exponent $\alpha$, 
the Brody parameter $\omega$, and the 
fraction $\rho_1^{cl}$ of regular classical trajectories 
of the Robnik billiard as functions of the deformation 
parameter $\lambda$.}
\label{fig:alpha_rho_omega} 
\end{figure}

Let us now compare the evolution of the parameter $\alpha$ as a
function of $\lambda$ with the fraction $\rho_1^{cl}$ of regular
classical trajectories in the phase space, and with the Brody
parameter $\omega$ \cite{Brody:81}. This is an {\em ad hoc} parameter,
without any known physical meaning, which quantifies to some extent the
chaoticity of the system. 
It nevertheless captures well the important feature of fractional 
power law level repulsion, that is the behavior of $P(s)$ at small $s$ 
\cite{Robnik:84, Prosen:93}. For $\omega=0$ we get the Poisson
distribution of regular systems, and at the other extreme, 
$\omega=1$, we obtain the Wigner distribution predicted by RMT for
chaotic systems.  Fig. \ref{fig:alpha_rho_omega} shows the behavior of
$\alpha$, $\omega$ and $\rho_1^{cl}$. There is a clear correlation
among these three variables, although the transition is smoother for
$\omega$ and especially for $\alpha$ than for $\rho_1^{cl}$.  However,
while the fraction of regular classical trajectories is almost zero
near $\lambda=0.25$, the power spectrum, and to a lesser extent the
$P(s)$ distribution, indicate an intermediate situation between
regular and chaotic motion. This clearly shows that $\omega$ and
$\alpha$ are not only functions of $\rho_1^{cl}$, but depend on finer
details of the underlying classical mechanics as well.

It is well known that in the strict semiclassical limit the quantum
eigenstates of a quantum system with generic (mixed) classical dynamics
can be classified as regular and irregular  following the
original proposition of Percival \cite{Percival:73}. 
This has been further developed and raised 
to a Principle of Uniform Semiclassical Condensation 
(PUSC) of Wigner functions of the eigenstates \cite{Berry:77b}. 
From this it follows that in the strict semiclassical limit 
the regular and irregular level sequences are statistically 
independent, but for themselves have Poisson or RMT level statistics, 
respectively. This theory has been excellently confirmed in hard numerical 
calculations \cite{Prosen:94}. 
Nevertheless, if the system is not deep enough 
in the semiclassical regime one can see substantial deviations from 
such a behavior, manifested in the fractional power law level 
repulsion \cite{Robnik:84, Prosen:93} 
in $P(s)$ at small $s$. A similar recent 
analysis \cite{Robnik:03} has demonstrated that in the strict 
semiclassical limit we should not expect a power law 
behavior for the power 
spectrum but something more complicated. Therefore it is 
quite an unexpected result of the present paper that 
the power spectrum $S(k)$ is a power law at all $k$ in 
mixed systems at low energies. Indeed, when going sufficiently 
deep into the semiclassical regime the theory of reference 
\cite{Robnik:03} should be expected and confirmed. 

In conclusion, the analogy between quantum energy spectra and time
series opens a new and fruitful perspective on the universal
properties of quantum level fluctuations. The $\delta_n$ function
gives the level fluctuations, and its power spectrum $S(k)$ is an
intrinsic characteristic of the quantum system. The important point is
that it exhibits a power law behavior, similar to the well known
$1/f^{\alpha}$ noise found in many complex systems. 

In the order to chaos transition, the chaoticity of a quantum system
is usually qualitatively assessed by how close to Poisson or RMT its
fluctuation properties are. In the present power spectrum approach,
the exponent changes smoothly from $\alpha=2$ for a regular system to
$\alpha=1$ for a chaotic system. Contrary to the Dyson $\Delta_3(L)$
statistic, that must be plotted for different values of $L$, the
exponent $\alpha$ quantifies the chaoticity of the system in a single
parameter. 
Moreover, $\alpha$ has a physical meaning. It is a natural measure of the 
fluctuation properties of a quantum system through the power spectrum, 
and provides an intrinsic quantitative measure of the regular and chaotic 
dynamical features.

The origin of the universal power law behavior $S(k) \sim
1/k^{\alpha}$ is now understood in the integrable case ($\alpha =2$)
and in the fully chaotic case ($\alpha =1$) on the basis of RMT
\cite{Faleiro:04} and semiclassical periodic orbit theory
\cite{Robnik:03}. The origin of the $1/f^{\alpha}$ power law in the
mixed regime still remains as an important open problem.

This work is supported in part by Spanish Government grants
BFM2003-04147-C02 and FTN2003-08337-C04-04.  This work is also
supported by the Ministry of Education, Science and Sports of the
Republic of Slovenia.

\end{document}